\documentclass{elsart}
\usepackage{natbib}
\usepackage{graphicx}
\usepackage{amssymb}

\begin{document}

\begin{frontmatter}
\title{Dictionary based methods for information extraction}

\author[fis]{A. Baronchelli},
\author[mat]{E. Caglioti},
\author[fis]{V. Loreto\corauthref{cor}}
\corauth[cor]{Corresponding author}
\ead{loreto@roma1.infn.it},
\author[iss]{E. Pizzi},
\address[fis]{``La Sapienza'' University, Physics Department, \\
  P.le A. Moro 5, 00185, Rome, Italy}
\address[mat]{``La Sapienza'' University, Mathematics Department, \\
  P.le A. Moro 2, 00198, Rome, Italy}
\address[iss]{Istituto Superiore di Sanit\`a \\
Viale Regina Elena 299, 00161 Rome, Italy}

\begin{abstract}
In this paper we present a general method for information extraction
that exploits the features of data compression techniques. We first
define and focus our attention on the so-called \textit{dictionary} of
a sequence. Dictionaries are intrinsically interesting and a study of
their features can be of great usefulness to investigate the
properties of the sequences they have been extracted from (e.g. DNA
strings). We then describe a procedure of string comparison between
dictionary-created sequences (or \textit{artificial texts}) that gives
very good results in several contexts. We finally present some results
on self-consistent classification problems.
\end{abstract}


\end{frontmatter}

\section{Introduction}
Strings or sequences of characters appear in almost all
sciences. Examples are written texts, DNA sequences, bits for the
storage and transmission of digital data etc. When analysing such
sequences the main point is extracting the information they bring. For
a DNA sequence this could help in identifying regions involved in
different functions (e.g. coding DNA, regulative regions, structurally
important domains) (for a recent review of computational methods in
this field see~\citep{Jiang}). On the other hand for a written text
one is interested in questions like recognizing the language in which
the text is written, its author or the subject treated.  \\ When
dealing with information related problems, the natural point of view
is that offered by Information Theory~\citep{shannon,zurek}. In this
context the word information acquires a precise meaning which can be
quantified by using the concept of entropy. Among several equivalent
definitions of entropy the best one, for our purposes, is that of
Algorithmic Complexity proposed by Chaitin, Kolmogorov and Solomonoff
~\citep{livit}: the Algorithmic Complexity of a string of characters
is the length, in bits, of the smallest program which produces as
output the string and stop afterward. \\ Though it is impossible, even
in principle, to find such a program, there are algorithms explicitly
conceived to approach such theoretical limit. These are the file
compressors or zippers. In this paper we shall investigate some
properties of a specific zipper, LZ77~\citep{LZ77}, used as a tool for
information extraction.

\section{The dictionary of a sequence}

It is useful to recall how LZ77 works. Let $x=x_1,....,x_N$ be the
sequence to be compressed, where $x_i$ represents a generic character
of sequence's alphabet. The LZ77 algorithm finds duplicated strings in
the input data. The second occurrence of a string is replaced by a
pointer to the previous string given by two numbers: a distance,
representing how far back into the window the sequence starts, and a
length, representing the number of characters for which the sequence
is identical. More specifically the algorithm proceeds sequentially
along the sequence. Let us suppose that the first $n$ characters have
been codified. Then the zipper looks for the largest integer $m$ such
that the string $x_{n+1},...,x_{n+m}$ already appeared in
$x_1,...,x_n$. Then it codifies the string found with a two-number
code composed by: the distance between the two strings and the length
$m$ of the string found. If the zipper does not find any match then it
codifies the first character to be zipped, $x_{n+1}$, with its
name. This eventuality happens for instance when codifying the first
characters of the sequence, but this event becomes very infrequent as
the zipping procedure goes on.  \\ This zipper has the following
remarkable property: if it encodes a text of length $L$ emitted by an
ergodic source whose entropy per character is $h$, then the length of
the zipped file divided by the length of the original file tends to
$h$ when the length of the text tends to
infinity~\citep{lz77optimal}. In other words LZ77 does not encode the
file in the best way but it does it better and better as the length of
the file increases. Usually, in commercial implementations of LZ77
(like for instance $gzip$), substitutions are made only if the two
identical sequences are not separated by more than a certain number
$n$ of characters, and the zipper is said to have a $n$-long sliding
window. The typical value of $n$ is 32768. The main reason for this
restriction is that the search in very large buffers could be not
efficient from the computational time point of view. A restriction is
often given on the length of a match, too, avoiding substitution of
repeated subsequences shorter than 3 characters.
\begin{figure}
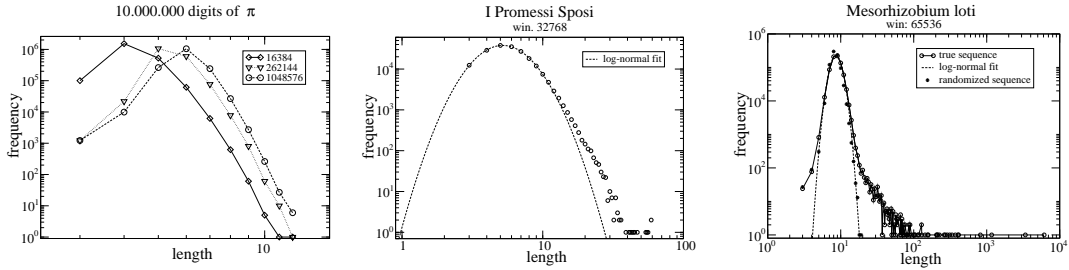

\begin{center}
\begin{tabular}{ccc}
{\includegraphics[height=3.5cm,angle=0]{pigTproc.eps}} 
&{\includegraphics[height=3.5cm,angle=0]{pspproc.eps}} 
&{\includegraphics[height=3.5cm,angle=0]{mslotiproc.eps}} 
\\
\end{tabular}
\caption{Frequency-Length distributions for words in the dictionaries
of different sequences. Left: The sequence of the first $10^6$
characters of $\pi$. Dictionaries extracted with different window
lengths. Center: The Italian book ``I Promessi Sposi'', with a
log-normal fit of the peak of the distribution. Right:
\textit{Mesorhizobium loti} original and reshuffled sequences, with
the log-normal fit of the peak.}
\end{center}
\vspace{0.cm}
\end{figure}
\\ We define \textit{dictionary} \citep{long} of a string the whole set
of sub-sequences that are substituted with a pointer by LZ77 and we
refer to these sub-sequences as dictionary's \textit{words}. From the
previous discussion it is clear that the same word can appear several
times in our dictionary (the multiplicity being limited by the length of
the sequence). Moreover, the structure of a dictionary is determined by
the size of the LZ77 sliding window. In particular, it has
been shown~\citep{lz77optimal,shannon-lec} that the average word length
$l$ found by an $n$-long sliding window LZ77 goes asymptotically as $l =
\frac{\log n}{h}$, where $h$ is the the entropy of
the (ergodic) source that emitted the sequence. It follows that the size
of the sliding window does not affect the number of characters in the
dictionary, but the way they are combined into words.  \\ In Figure
1 the frequency-length distributions for the words in the dictionaries
of several sequences of increasing complexity are presented. In each
figure the number of words of any length is plotted. For the sequence of
digits of $\pi$ (which can be assumed to be a sequence of
realizations of independent and identically distributed random variables) the spectra
obtained for three different sizes of the LZ77 sliding window are
presented. As expected the peak of the distribution grows with the
window's size. In the central plot the dictionary of the Italian book
``I Promessi Sposi'' is analysed. In this case, while the peak is well
fitted by a log-normal distribution (i.e. a Gaussian in logarithmic
scale), several very long words appear. The presence of long words
becomes crucial in the dictionary extracted by the DNA sequence of
\textit{Mesorhizobium loti} in the right plot. Here we compare the
dictionary extracted from the true sequence with the one obtained from
its randomization. As expected, long words are absent in the
dictionary of the reshuffled sequence.  \\ 
Since a genome is composed of regions coding for proteins (genes) and
of intergenic non-coding tracts, we have analysed in more detail the
contribution of these parts to the distributions of repeated ``words''. In
Figure 2 results obtained in the case of \textit{Escherichia coli} genome are
reported.  This genome is approximately 4.500.000 base pairs long; the
$87\%$ belongs to coding regions (see dotted line in the figure on the
right). In the figure on the left, the frequency-length  
distributions for the entire genome and for the coding tracts are
reported. The two distributions appear as completely overlapped up
to 20 base pairs of length, while for the next lengths they deviate from
each other. This fact is highlighted in the figure on the left, where
the fraction of words of each length coming from coding regions is
reported. It is clearly visible that within a range of
approximately 20 - 90 base pairs, most words come from non-coding
tracts. We observed an analogous behavior in the \textit{Vibrio cholerae}
second chromosome analysis (data not shown).   
It is a well known fact that non-coding sequences are characterized by
the presence of repeated ``words", however, at least for the analysed
prokaryotic genomes, our results seem to suggest that these tracts are
not more repetitive than genes but, more precisely, that they are
characterized by repeated words longer than those occurring within
coding parts.  Furthermore these preliminary results suggest our
approach as an useful tool to study genomes and their organization. 
%The presence of long repeated sequences in DNA strings is
%a well known fact in biology. In Figure 2 the results of a more detailed
%investigation of this phenomenon are shown. We have determined which
%fraction of words of each length comes from a coding region (i.e. from a
%region which contains information for the production of proteins). In
%particular the figure shows data for Escherichia coli. This bacterium
%has a DNA of approximately 4.500.000 basis, $87\%$ of which belongs to
%coding regions. In figure is clearly visible a range of lengths
%(approximately 20-90 basis) for which most words come from non coding
%regions. We observed a qualitative analogous behavior in the Vibrio
%cholerae second chromosome analysis (data not shown). We think that this
%result, if confirmed in more detailed analysis, can be important in
%the context of the studies of statistical properties of DNA sequences.
\begin{figure}
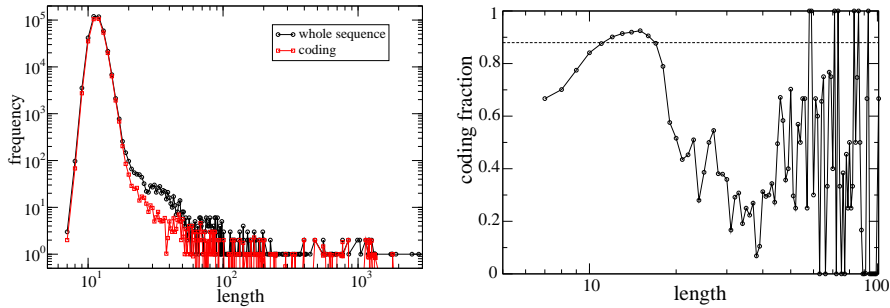

\begin{center}
\begin{tabular}{cc}
{\includegraphics[height=4cm,angle=0]{Ecolilufrproc2.eps}} 
&{\includegraphics[height=4cm,angle=0]{Ecolifrazproc.eps}} 
\\
\end{tabular}
\caption{Fraction of words extracted from coding regions
(\textit{Escherichia coli}). This
dictionaries were extracted giving LZ77 the possibility of finding
repeated sequences in the whole string. Words of lengths between 20
and 90 characters are found to belong mainly to non-coding regions.} 
\end{center}
\vspace{0.cm}
\end{figure}

\section{Dictionary-based self classification of corpora}

Data compression schemes can be also used to compare different
sequences.  In fact it has been shown \citep{loewenstern, khmelev,
bcl} that, compressing with LZ77 a file B appended to a file A, it is
possible to define a remoteness between the two files.  More precisely
the difference between the length of the compressed file A+B and the
length of the compressed file A, all divided by the length of the file
B, can be related to the cross entropy\footnote{With the term
cross-entropy between two strings we shall always refer in this paper
to an estimate of the true cross-entropy between the two ergodic
sources from which A and B have been generated.} between the two files
\citep{paper-yak}.This method is strictly related to the algorithm by
Ziv and Merhav \citep{ziv-merhav} which allows to obtain a rigorous
estimate of the cross entropy between two files A and B by
compressing, with an algorithm very similar to LZ77, the file B in
terms of the file A.  In \citep{bcl} experiments of language
recognition, authorship attribution and language classification are
performed exploiting the commercial zipper \textit{gzip} to implement
the technique just discussed. In this paper we use a natural extension
of the method used in~\citep{bcl}, devised to measure directly the
cross entropy between A and B: in particular the LZ77 algorithm only
scans the B part and looks for matches only in the A part. In
experiments of features recognition (for instance language or
authorship) a text X is compared with each text $A_i$ of a corpus of
known texts. The closest $A_i$ sets the feature of the X text
(i.e. its language or author). In classification experiments, on the
other hand, one has no a priori knowledge of any texts and the
classification is achieved by the construction of a matrix of the
distances between pairs of sequences. A suitable tree representation
of this matrix can be obtained using techniques mutuated from
phylogenetics.  It must be underlined that, for self-consistent
classification problems, a true mathematical distance is needed (see
for a discussion~\citep{livit,bennett,bcl}). 

\begin{figure}
\begin{center}
\begin{tabular}{cc}
{\includegraphics[height=7cm,angle=0]{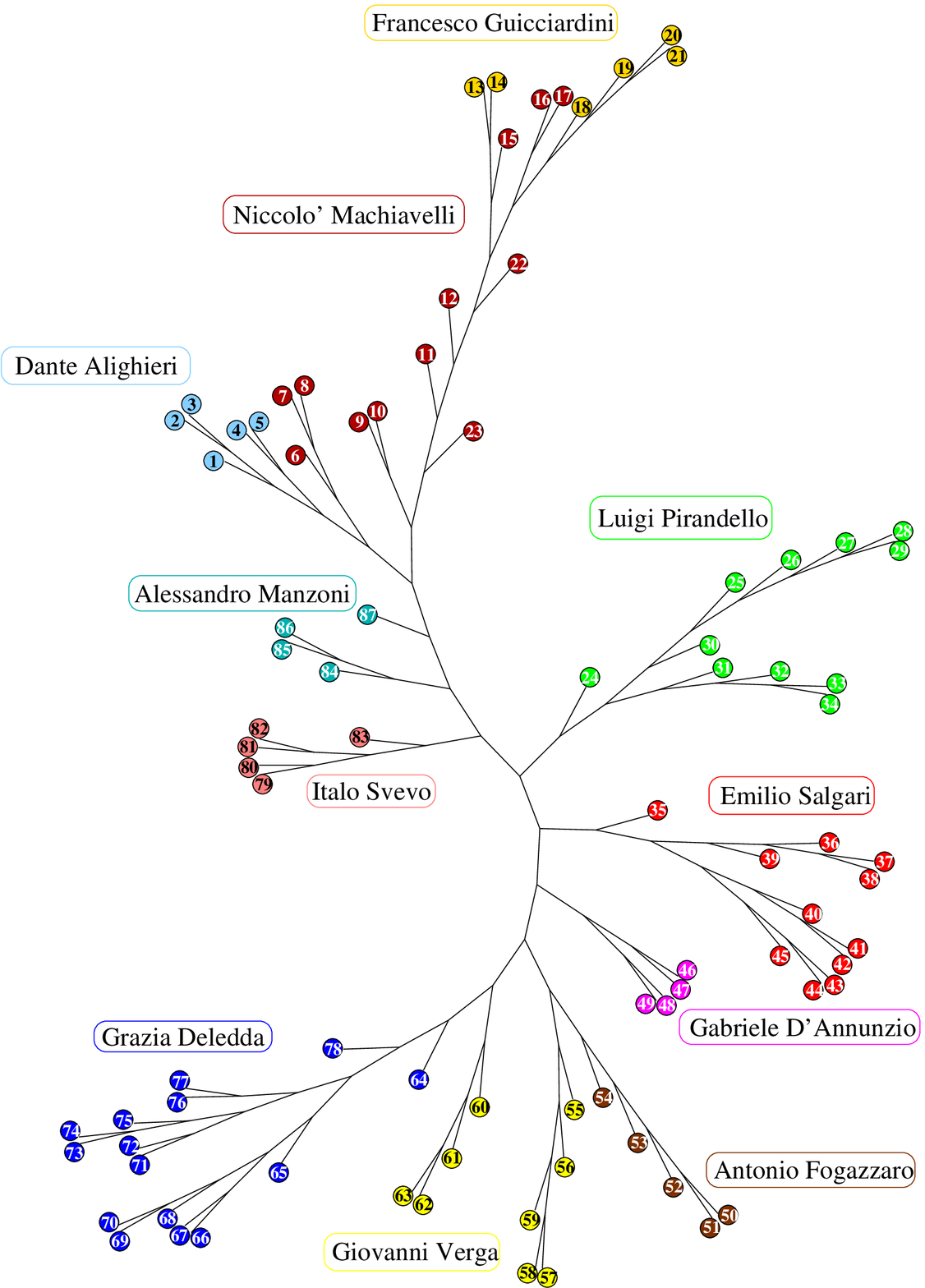}} 
&{\includegraphics[height=7cm,angle=0]{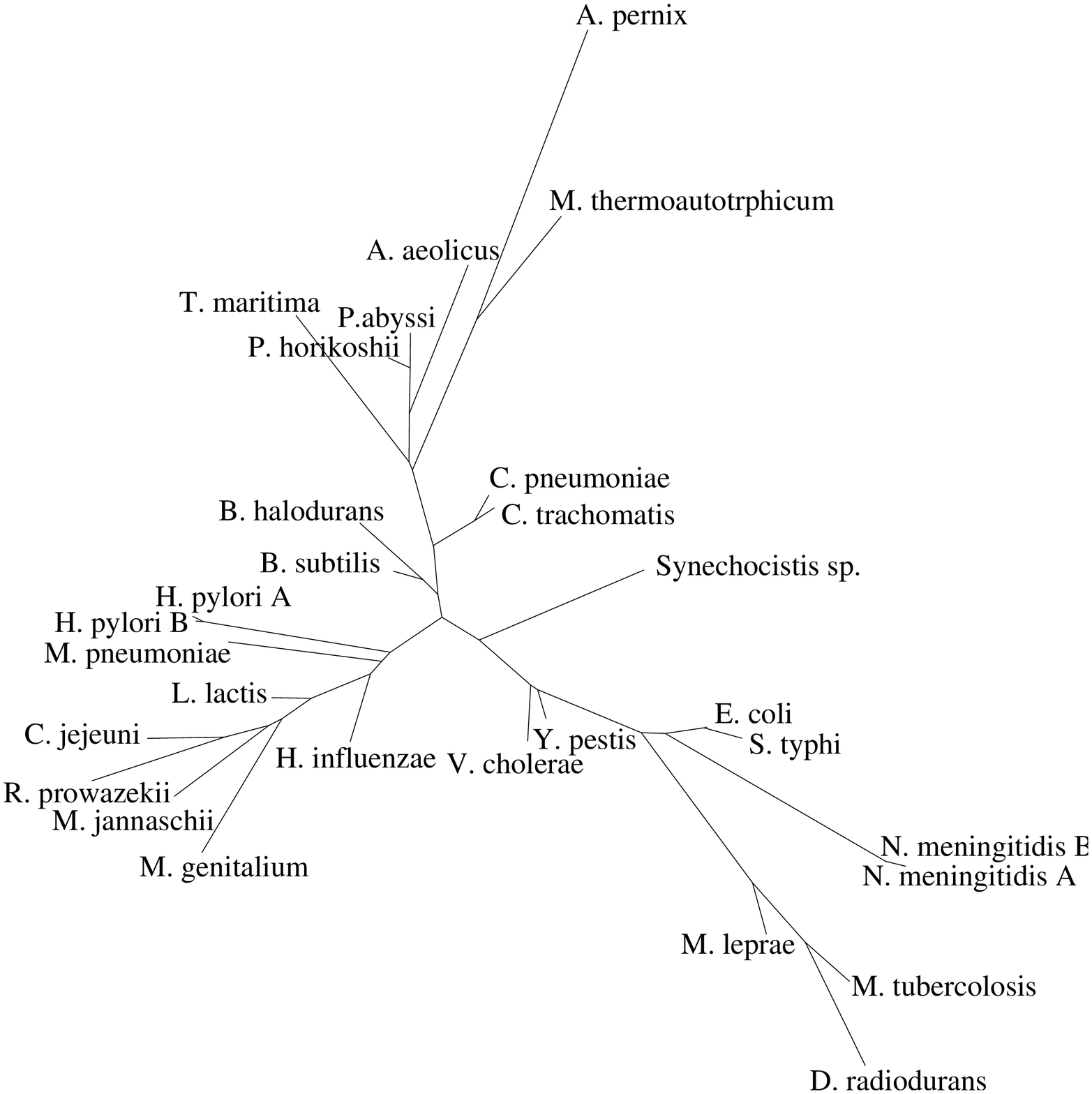}} 
\\
\end{tabular}
\end{center}
\caption{Self-consistent classification. Left: a tree obtained from a
corpus of 87 works of 11 Italian writers. Right: a species similarity
tree for 27 procariotes. Both trees have been obtained from distance
matrices constructed with the artificial texts comparison method.} 
\end{figure}

Our idea (see also \citep{long}) is that of creating
\textit{artificial texts} by appending words randomly extracted from a
dictionary and to compare artificial texts instead of the original
sequences. The comparison of artificial texts is made using the
modified version of LZ77 discussed above. One of the biggest
advantages of our artificial text method is the possibility of
creating an ensemble of artificial texts all representing the same
original sequence, thus enlarging the original set of
sequences. Comparing artificial texts we performed the same
experiments described in \citep{bcl} obtaining better results.

In Figure 3 we present a linguistic tree representing the
self-classification of a corpus of 87 texts belonging to 11 Italian
authors \citep{liberliber}. The texts belonging to the same author
clusterize quite well, with the easily-explainable exception of the
Machiavelli and Guicciardini clusters. The other tree presented in
Figure 3 is obtained by a whole-genome comparison of 27 prokaryotic
genomes. This kind of analysis are now definitely possible thanks to
the availability of completely sequenced genomes (See for a similar
approach~\citep{mingli}).  Our results appear as comparable with those
obtained through other completely different "whole-genome" analysis
(see, for instance, \citep{pride}). Closely related species are
correctly grouped (as in the case of \textit{E.coli} and
\textit{S.typhimurium}, \textit{C.pneumoniae} and
\textit{C. trachomatis}, \textit{P. abyssi} and
\textit{P. horikoshii}, etc), and some main groups of organisms are
identified. It is known that the mono-nucleotide composition is a
specie-specific property for a genome. This compositional property
could affect our method: namely two genomes could appear as similar
simply because of their similar C+G content. In order to rule out this
hypothesis we performed a new analysis after shuffling genomic
sequences and we noticed that the resulting new tree was completely
different with respect to the one based on real sequences.

%In biology it is known that the GC
%content of a DNA sequence (i.e. the percentage of GC basis) is a
%signature of species. This content afflict the results of our method,
%too. To be sure it was not the leading element in our analysis, we
%constructed a tree starting from reshuffled sequences (data not shown)
%and the result was absolutely different (containing almost no
%information). Finally, it must be noted that our trees represent only
%similarity properties of different sequences, and can not be used to
%infer anything about the evolution of those sequences from a common
%ancestor. For this reason, strictly speaking, our trees can not be
%said to be phylogenetic. 

In conclusion we have defined the dictionary of a sequence and we have
shown how it can be helpful for information extraction
purposes. Dictionaries are intrinsically interesting and a statistical
study of their properties can be a useful tool to investigate the
strings they have been extracted from. In particular new results
regarding the statistical study of DNA sequences have been presented
here. On the other hand, 
we have proposed an integration of the string comparison procedure
presented in~\citep{bcl} that exploits dictionaries by means of
artificial texts. This method gives very good results in several
contexts and we have focused here on self-classification problems,
showing two similarity trees for corpora of written texts and DNA
sequences.

\end{document}